\documentclass[preprint,showpacs,amsmath,preprintnumbers,bibnotes,amssymb,aps,prd]{revtex4-1}

\usepackage{graphicx}
\usepackage{subfigure}
\usepackage{dcolumn}
\usepackage{bm}
\usepackage{amssymb}

\begin{document}
\title{Pulsar Timing Sensitivity to \\
  Very-Low-Frequency Gravitational Waves}

\author{Fredrick A. Jenet} 
\email[]{merlyn@phys.utb.edu }
\affiliation{Center for Gravitational Wave Astronomy, University of
  Texas, Brownsville TX 78520}

\author{J.W. Armstrong} 
\email[]{john.w.armstrong@jpl.nasa.gov}
\affiliation{Jet Propulsion Laboratory, California Institute of
  Technology, Pasadena CA 91109}

\author{Massimo Tinto} 
\email[]{massimo.tinto@jpl.nasa.gov}
\affiliation{Jet Propulsion Laboratory, California Institute of
  Technology, Pasadena CA 91109}

\date{\today}

\begin{abstract}
  At ~nanohertz frequencies gravitational waves (GWs) cause variations
  in time-of-arrival of pulsar signals potentially measurable via
  precision timing observations. Here we compute very-low-frequency GW
  sensitivity constrained by instrumental, propagation, and other
  noises fundamentally limiting pulsar timing observations.  Reaching
  expected GW signal strengths will require estimation and removal of
  $\simeq$99\% of time-of-arrival fluctuations caused by typical
  interstellar plasma turbulence and a reduction of white rms timing
  noise to $\sim$100 nsec or less.  If these were achieved,
  single-pulsar signal-to-noise ratio (SNR) = 1 sensitivity is then
  limited by the best current terrestrial time standards at $h_{rms}
  \sim$2 $\times 10^{-16}$ [f/(1 cycle/year)]$^{-1/2}$ for $f < 3
  \times 10^{-8}$ Hz, where f is Fourier frequency and a bandwidth of
  1cycle/(10 years) is assumed.  This sensitivity envelope may be
  optimistic in that it assumes negligible intrinsic pulsar rotational
  noise, perfect time transfer from time standard to observatory, and
  stable pulse profiles.  Nonetheless it can be compared to predicted
  signal levels for a broadband astrophysical GW background from
  supermassive black hole binaries.  Such a background is comparable
  to timekeeping-noise only for frequencies lower than about 1
  cycle/(10 years), indicating that reliable detections will require
  substantial improvements in signal-to-noise ratio through pulsar
  array signal processing.
\end{abstract}

\pacs{98.80.-k, 95.36.+x, 95.30.Sf}
\maketitle

The pulsar timing GW detector uses the earth and a distant pulsar as
electromagnetically-tracked separated test masses.  The pulsar
emission serves as a clock, in the idealized case producing perfectly
periodic radio pulses transmitted to the earth.  These are ÒtimedÓ by
cross correlation of the received pulses against a template of the
pulsed waveform.  Time-of-arrival residuals, R(t), are produced after
correcting for known effects.  Signals and noises enter the observed
time series via transfer functions \cite{EW-75}.  A GW of
characteristic strain amplitude h incident on the earth-pulsar system
produces variations of order h in the time series of relative
dimensionless frequency fluctuations, $y(t)$, of the pulsar signal
\cite{AET-99} $y(t) = ((1-\mu)/2) [\Psi(t - T_1(1 + \mu)) - \Psi(t)]$.
Here $\mu = \bf{k} \cdot \bf{n}$, $\bf{k}$ is a unit vector parallel
to the GW propagation direction, $\bf{n}$ is a unit vector from the
earth to the pulsar, $\bf{h}(t)$ is the GW strain tensor, $\Psi(t) =
(\bf{n} \cdot \bf{h} \cdot \bf{n})/(1 - \mu^2)$, and $T_1$ is the
one-way light travel time between the pulsar and earth.  The
fractional frequency time series is the derivative \cite{detweiler-79}
of the observed time-of-arrival residuals, R(t): y(t) = $dR(t)/dt$.
The pulsar timing technique has been used to bound GW signal
strengths, e.g.
\cite{HD-83,RT-83,KTR-94,JLLW-04,JHLM-05,VBCHVCJMBSYBHY-09}.

To assess instrumental and other noises currently and fundamentally
limiting detections, we compute here the sensitivity of pulsar GW
observations.  Sensitivity is conventionally expressed as the sky- and
polarization-averaged sinusoidal signal strength necessary to achieve
a given signal-to-noise ratio in a given bandwidth, e. g.
\cite{AET-99,armstrong-06}.  Explicitly, we compute the signal
strength required to produce SNR = 1 in bandwidth B: $[S_{yn}(f)
B]^{1/2}$/(rms signal response), where $S_{yn}(f)$ is the spectrum of
noise and the rms signal response in general also depends on Fourier
frequency.

The GW signal response depends on the angle of arrival of a wave
relative to the earth-pulsar line.  For GWs from a specific direction
the above formula for y(t) can be used directly \cite{JLLW-04}.  We
are interested here in signal response averaged over the sky.  To get
the rms signal response as a function of Fourier frequency, the
Fourier transform squared of the GW signal response, above, is
averaged over the sky and polarization states to obtain
$S_{y}(f)/S_h(f) = 1/3 - 1/(8\pi^2 f^2 {T_1}^2) + \sin(4 \pi f
T_1)/(32 f^3 \pi^3 {T_1}^3)$, where $S_y$ is the spectrum of
fractional frequency fluctuations, $S_h$ is the spectrum of GW
strengths, and f is Fourier frequency.  In the practical case $T_1$ is
hundreds of years or longer and the duration of pulsar timing
observations is $\sim$decades, so the second and third terms are
negligible for $f > 1$/(duration of observations): $S_y(f)/S_h(f)
\approx 1/3$, implying that rms signal response is constant ($\simeq
0.58$) over the accessible frequency band.

\begin{figure}
\centering
\includegraphics[width=6.25in]{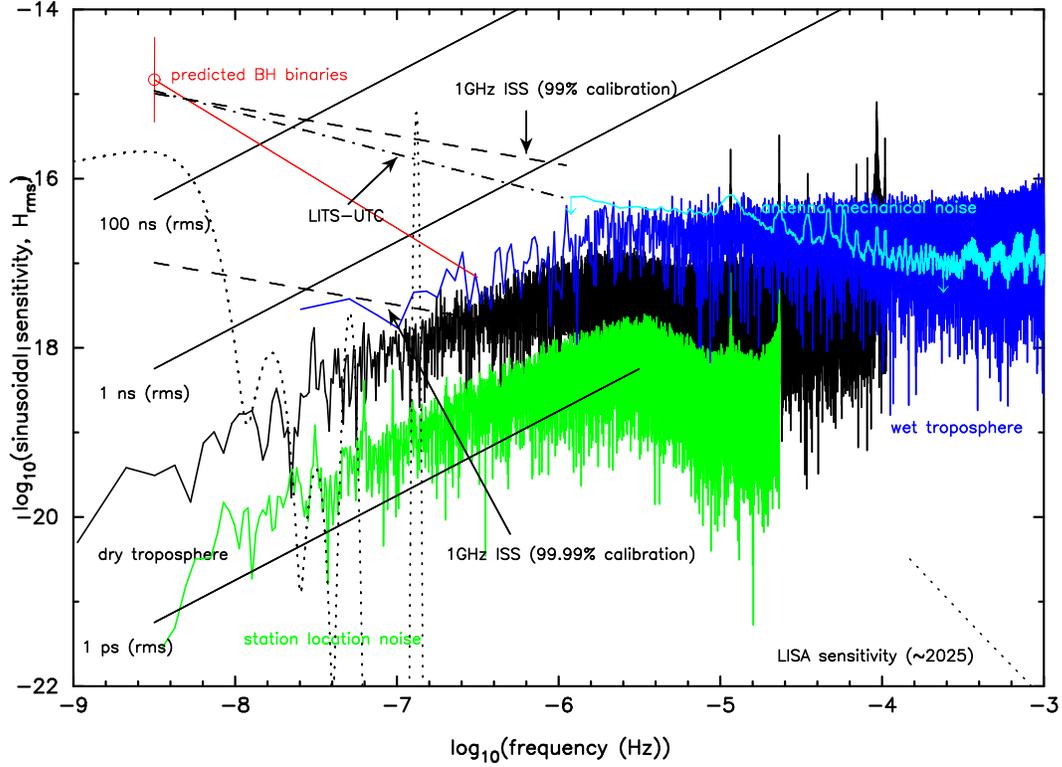}
\caption{Gravitational wave sensitivity expressed as strain amplitude
  required for SNR = 1 in a 1 cycle per 10 years bandwidth, as a
  function of Fourier frequency.  Sensitivity limited by various
  noises are indicated: green is station location noise \cite{aplo};
  black and blue are respectively due to fluctuations in the zenith
  dry \cite{noaa} and wet \cite{keihm-95} troposphere; light blue is
  due to antenna mechanical noise for a 34m beam-waveguide
  station\cite{AITB-03,armstrong-06}; dashed lines are for dispersion
  measure variations in the interstellar medium
  \cite{armstrong-84,ARS-95} for 1 GHz observations after $99\%$ and
  $99.99\%$ calibration; solid black lines are for white timing noise
  with rms amplitudes of 100 ns, 1ns, and 1ps in a Fourier band
  $\pm0.5$ cycles/day; dotted line is an approximate limit due to
  uncertainties in the masses of the planets
  \cite{lommen-01,standish-07}; dot-dashed line is sensitivity limited
  by a linear ion trap time standard \cite{BDT-08}.  Also shown for
  reference is the low-frequency sensitivity expected for the LISA
  detector \cite{ETA-00}.  Red line and vertical bar shows the
  dependence and range of predicted signal strengths from an ensemble
  of supermassive black-hole binaries \cite{RR-95,JB-03,WL-03,EINS-04}
}
\label{Fig1}
\end{figure}

To compute sensitivity we also need spectra of the noises.  Important
noise sources include finite signal-to-noise ratio in the raw
observations, instability of the local clock against which pulsars are
timed (and errors in time transfer if the clock is not located at the
observatory), uncertainties in solar system ephemerides (used to
correct arrival times at the earth to the barycenter of the solar
system), pulsar position uncertainty, intrinsic pulsar rotational
stability, stability and accuracy of the average pulse templates used
to measure R(t), dispersion measure (DM) variability in the
interstellar and interplanetary plasmas, tropospheric scintillation
(the wet and dry components of the troposphere cause delay
variability), antenna mechanical noise (stability of the phase center
of the antenna tracking the pulsar), and station location errors
(changes in antenna location due to atmospheric and tidal loading of
the crust).  Figure (\ref{Fig1}) shows the GW sensitivity for several
of these noise sources individually, as discussed briefly below, with
sensitivity computed for bandwidth B = 1 cycle/10 years.

The green curve is the station-location-noise limit; $S_{yn}$ was
computed from 30 years of absolute value of vector ground displacement
using data \cite{aplo} taken near the NASA/JPL Goldstone CA tracking
complex.  The derivative theorem for Fourier transforms
\cite{bracewell-65} was used to convert the spectrum of displacement
to the spectrum of velocity and hence the spectrum of y = $\Delta$v/c.
The black curve is derived from the power spectrum of hourly zenith
dry tropospheric pressure fluctuations \cite{noaa} provided by the
National Climate Data Center and taken at the NASA/JPL deep space
tracking complex near Madrid, Spain.  Pressure was converted to zenith
path variation using (path variation in centimeters) =
0.022768*(surface pressure in millibars), ignoring a factor close to
unity which depends on latitude and height.  The blue curve is from
the power spectrum of zenith wet tropospheric path delay, computed
from 1.5 years of data \cite{keihm-95} taken at the NASA/JPL
Goldstone, California tracking complex.  Tropospheric path variation
spectra were similarly converted to spectra of y using the derivative
theorem.  The light blue curve is the measurement (for f $>$ 0.0001
Hz) and upper limit (for $10^{-6} - 10^{-4}$ Hz) for antenna
mechanical noise fluctuations observed with a 34m tracking antenna at
Goldstone \cite{AITB-03}; smaller stiffer antennas give lower antenna
mechanical noise \cite{ALMA_antennas1} \cite{MBGLSWH-06}.  The solid
black lines are for white timing noise with rms amplitudes of 100
nsec, 1 nsec, and 1 ps in a Fourier band $\pm$ 0.5 cycles/day (i.e.
one sample per day; current observations are more typically 1 sample
per 2 weeks, which would result in curves $\sqrt{14}$ higher).  The
100 nsec level is the current timing goal of leading timing array
experiments; three pulsars are being timed to this level
\cite{VBCHVCJMBSYBHY-09}.  One picosecond is the absolute best
possible timing accuracy one can achieve using millisecond pulsars.
Since the pulsar signal itself is an amplitude modulated noise
process, it can be said to have Ôself-noiseÕ.  In the absence of all
other sources of noise, including timing and antenna noise, the pulsar
signal self-noise would still be present.  Assuming that the narrowest
possible average pulse profile is 10 $\mu$sec, the pulsar signal
bandwidth is 1 GHz, and that one can observe the pulsar for 12 hours
at a time, the self-noise yields an rms timing accuracy of about 1 ps.
The dotted curve shows approximate limits due to the uncertainties in
the masses of the planets
\cite{lommen-01,standish-07,planetary_mass-10}, Mercury through
Jupiter, affecting knowledge of the solar system barycenter.

The dashed lines are noise from representative interstellar medium
dispersion measure (DM) variation \cite{armstrong-84,ARS-95} (assumed
Kolmogorov spectrum with ${C_n}^2 = 0.001$ m$^{-20/3}$, propagation
distance z = 1 kpc, radio frequency = 1 GHz, transverse velocity v =
100 km/sec), $S_{yn}(f) = \pi^{-1/6} 2^{-2/3} v^{5/3} \lambda^4 c^{-2}
z {C_n}^2 {r_e}^2 [\Gamma[4/3]/\Gamma[11/6]] f^{-2/3}$, with the
indicated levels of calibration, i.e. $99\%$ calibration means only
$1\%$ of the DM fluctuation rms noise remains in the measurement.
(Some nearby pulsars have smaller integrated ISM turbulence levels and
would require smaller percentage corrections to reach the indicated
line in Figure (\ref{Fig1})).  The effect of solar wind plasma
turbulence is non-negligible \cite{YHCMH-07} but its dispersive
character should allow it to be calibrated in addition to the
interstellar plasma.

Pulsars are timed against terrestrial clocks.  Recent stability
measurements of linear ion trap time standards \cite{BDT-08} give
$S_{yn}(f) \simeq 4 \times 10^{-31} (f/1Hz)^{-1}$, measured in the
approximate band $10^{-7} - 10^{-6}$ Hz.  We assume this spectrum
continues to be valid to lower frequencies; in a 1 cycle/(10 year)
bandwidth this noise gives GW sensitivity shown as the dot-dash line
in Figure (\ref{Fig1}).  The dotted curve in the lower right, for
comparison, is the low-frequency segment of the Laser Interferometer
Space Antenna (LISA) missionÕs predicted sensitivity curve
(\cite{ETA-00}, 5-$\sigma$-in-one-year-integration).

Not included in Figure (\ref{Fig1}) are intrinsic pulsar rotational
instability noise \cite{LK-05,SC-10} (variable by pulsar and
substantial for some), errors due to time transfer from the frequency
standard to the observatory \cite{levine-05}, pulsar position
uncertainty \cite{LK-05}, errors in pulse templates, and errors due to
radio frequency dependence and temporal instability of pulse profiles.
So the upper envelope sensitivity in Figure (\ref{Fig1}) is in this
sense optimistic.

One application of our sensitivity analysis is the detectability of an
astrophysical GW background from incoherently radiating supermassive
black hole binaries.  Such a background is predicted to produce signal
strengths \cite{RR-95,JB-03,WL-03,EINS-04,VBCHVCJMBSYBHY-09} in the
range $h(f) = [f S_h(f)]^{1/2} \sim (1 - $to$ - 10) \times 10^{-16}
$[f/(1 cycle/year)]$^{-2/3}$.  For comparison with our SNR = 1 in a
fixed B = 1 cycle/(10 years) bandwidth, we convert to $h_{rms}(f) = [B
S_h(f)]^{1/2} \sim (3-$to$-30) \times 10^{-17}$ [f/(1
cycle/year)]$^{-7/6}$.  These GW strengths are comparable to the SNR =
1 sensitivity limited at low frequencies by time-standard noise only
for $f \sim$ 1 cycle/(10 years) or lower (Figure (\ref{Fig1})).
(Within a broadband astrophysical background there may be some sources
strong enough to be detected individually, i.e. detectable GWs coming
from specific directions.  Figure (\ref{Fig1}) shows how strong an
individual source would have to be for detection above the timekeeping
noise limit.  When simultaneously timing several pulsars, the
directional property of the timing response to gravitational radiation
from a single source can be used in the same way as for ground-based
networks of broadband GWs to improve SNR, e.g.  \cite{SST-09} and
references therein.)  Since SNR $>$ 5 is conventionally taken as
detection threshold, Figure (\ref{Fig1}) indicates that substantial
SNR improvements will be required of pulsar timing array signal
processing for reliable detections.

\begin{acknowledgments}
  We thank F. B. Estabrook for discussions on gravitational wave
  sensitivity and W. A. Coles for comments on an early draft of this
  paper.  F. A. J.'s contribution was funded by a grant from the U. S.
  National Science Foundation (AST \#0545837).  For J. W. A. and M.
  T., this research was carried out at the Jet Propulsion Laboratory,
  California Institute of Technology, under a contract with the
  National Aeronautics and Space Administration and funded through the
  internal Research and Technology Development program.
\end{acknowledgments}

\bibliography{prd_version}
\end{document}